\documentclass{emulateapj}

\begin{document}

Astrophysical Journal Letters, Accepted

\newcommand{\mjybm}{\mbox{mJy~beam${}^{-1}$}}
\newcommand{\sgrastar}{\protect\objectname[]{Sgr~A$^{*}$}}
\newcommand{\htwo}{\ion{H}{2}}

\shorttitle{\sgrastar\ at~330~MHz}
\shortauthors{Nord et al.}

\title{Detection of Sagittarius~A$^*$ at~330~MHz with the Very Large Array}

\author{Michael E.~Nord\altaffilmark{1}, T.~Joseph~W.~Lazio, Namir~E.~Kassim}
\affil{Naval Research Laboratory, Code~7213, Washington, DC  20375-5351}
\email{Michael.Nord@nrl.navy.mil}
\email{Joseph.Lazio@nrl.navy.mil}
\email{Namir.Kassim@nrl.navy.mil}
\altaffiltext{1}{Department of Physics and Astronomy, University of New Mexico}

\author{W.~M.~Goss}
\affil{National Radio Astronomy Observatory, P.O.~Box~0, Socorro, NM 87801}
\email{mgoss@nrao.edu}

\and
\author{N.~Duric}
\affil{Department of Physics and Astronomy, University of New Mexico, 800 Yale 
Boulevard NE, Albuquerque, NM  87131}
\email{duric@tesla.phys.unm.edu}

\begin{abstract}

We report the detection of Sagittarius~A$^*$, the radio source associated with our 
Galaxy's central massive black hole, at 330 MHz with the Very Large Array.  Implications for the 
spectrum and emission processes of Sagittarius~A$^*$ are discussed and several hypothetical geometries of the central region are considered.


\end{abstract}


\section{Introduction}\label{sec:intro}

The central radio-bright region of our Galaxy, known as the Sagittarius Complex,  
is composed of three major components:  The supernova remnant (SNR) Sagittarius 
A East (\objectname[]{Sgr~A East}), the Sagittarius A West (\objectname[]{Sgr~A 
West}) \htwo\ region, and Sagittarius A$^*$ (\sgrastar), recently established as 
our Galaxy's central massive black hole (e.g., Ghez et al.~2000; Eckart et 
al.~2002).

Models attempting to explain the emission from \sgrastar\ fall into three broad 
classes.  Emission is modeled as arising from thermal sources, such as a low-temperature accretion 
disk, from non-thermal sources such as a jet (e.g., Melia \& Falcke, 2001 and references therein), or from a mixture of the two.  Such models are constrained primarily by the observed spectrum of \sgrastar, but large gaps in frequency coverage exist.  For this reason, filling in such gaps, as with the recent near-IR detections (Ghez et al.~2003, submitted; Genzel et al.~2003), and extending the range of frequencies over which the source is observed is important in order to place additional observational constraints on these models.

In 1976, Davies et al.~attempted to observe \sgrastar\ at 410 MHz, but reported 
no detection above a level of 50 mJy.
The first Very Large Array (VLA) images of the Galactic Center (GC) at 330~MHz 
(Pedlar et al.~1989, Anantharamaiah et al.~1991) represented a major improvement 
in sensitivity and resolution over previous meter wavelength images.  Pedlar et 
al.~did not detect \sgrastar\ down to a 5$\sigma$ level of 100 mJy. Though 
Anantharamaiah et al.~had higher sensitivity, (rms $\sim 3$ \mjybm) they also 
reported a non-detection.  The absence of any detections below 1000 MHz was 
interpreted as thermal absorption obscuring \sgrastar, indicating the low 
frequency properties of \sgrastar\ are dominated by extrinsic absorption, not 
intrinsic emission. 

New low frequency detections of \sgrastar, including the first detection of the source below 1 GHz 
(610 MHz; Roy et al.~2003), and this 330 MHz detection suggest that the low frequency 
properties of \sgrastar\ and the implied line of sight optical depths need to be
re-examined.

\section{Observations and Data Reduction}\label{sec:observations}

The Galactic Center was observed at~330~MHz using the Very Large Array
in its A configuration; Table~\ref{table:observe} summarizes
the observations.  The visibility amplitudes were calibrated with
observations of \objectname[3C]{3C~286}, while the visibility phases
were calibrated with observations of the VLA calibrators
\objectname[VLA]{J1751$-$253} and \objectname[VLA]{1714$-$252}.

\begin{deluxetable}{lcccc}
\tablecaption{Observational Summary}
\tablehead{\colhead{Epoch} & \colhead{Config.} & \colhead{$\nu$ (MHz)} & 
\colhead{$\Delta\nu$(MHz)}& \colhead{Integration(hours)} }
\startdata

1996 Oct.  & A & 332.5 & 6 & 5.6 \\
1998 Mar.  & A & 327.5 & 3 & 5.4 \\

\enddata
\label{table:observe}
\end{deluxetable}

In this letter we will summarize only salient details of the image processing.
Full details of the processing including further details on issues such as radio frequency interference excision, astrometry, bandwidth smearing, and extensive details pertaining to calibration and ionospheric compensation will be provided in \cite{nordetal03}.

Observations were taken in 
spectral-line mode, both to reduce the impact of bandwidth smearing
and to enable excision of radio frequency interference.  Excision was done 
manually by searching for visibilities
with anomalously high amplitudes, relative to surrounding visibilities
in the $u$-$v$ plane as well as by identifying the Fourier components
that gave rise to systematic ``ripples'' in the image.  In general,
low-frequency observations with the VLA must take the non-coplanarity
of the array into account.  Here this was not done because the field of 
view required to image Sgr A East and West is sufficiently limited ($\approx 
3\arcmin$) that the array can be assumed to be nearly coplanar.  Several
iterations of phase-only self-calibration followed by
imaging were performed in order to improve the dynamic range of the
images and compensate for ionospheric-induced phase distortions.
Because data are being combined from different epochs, each epoch was imaged separately to
produce an image with an acceptable dynamic range before being combined. The 
resulting self-calibrated $u$-$v$ data from the separate epochs were combined 
and phase and amplitude calibrated against the 1998 image in order 
to account for any offset in flux density scales and coordinates grids.

\begin{figure}[tbp]
\plotone{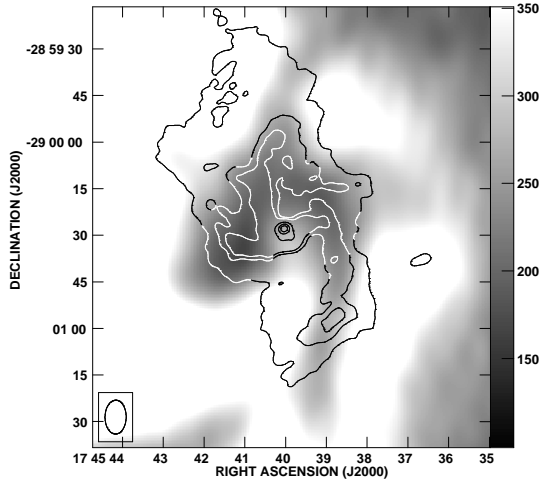}
\caption{Super-uniform image of the \sgrastar\ region at 330~MHz.
White areas represent areas of high brightness, dark areas 
represent areas of lower brightness due to absorption of Sgr A East SNR by Sgr A 
West mini-spiral and Sgr A West (extended).  The grey scale is linear 
between~$100$ and 350 \mjybm\arcdeg.  The resolution is $6\farcs8 \times 10\farcs9$ 
with a position angle of $-1\arcdeg$ and an rms noise of $\sim$ 4 \mjybm.  Contours are
5 GHz (Yusef-Zadeh \& Morris~1987) with a resolution of $3\farcs4 \times 2\farcs 9$, position angle of $-80\arcdeg$, and
levels of 1, 2, 3, 9, 15, 20, 25, and 30 times 40 \mjybm.  \sgrastar is located at the position of maximum 5~GHz intensity.  Note the 330 MHz absorption
and the 5 GHz diffuse flux density correlate well except at the position of \sgrastar\ where the
5~GHz is maximum and the 330~MHz is at a local maximum.}
\label{fig:sgrastar}
\end{figure}

In order to obtain maximum resolution, the super-uniform weighting scheme 
\citep{1999sira.conf..127B} was used.  Super-uniform 
weighting\footnote{\texttt{UVBOX$=3$} and \texttt{ROBUST$=-1$} within the AIPS 
task \texttt{IMAGR}.} reduces the weight, both globally and locally, given to 
the central, densely
sampled region of the $u$-$v$ plane, maximizing the resolution at the expense of 
increased noise and sidelobe level.   In this case the increase in noise and 
sidelobe level is an acceptable trade off, as the analysis is limited more by
the $u$-$v$ coverage and resulting image fidelity than by the thermal
noise level.  Figure~\ref{fig:sgrastar} shows the super-uniform image of the 
\sgrastar\ region.

\section{Analysis}\label{sec:analysis}

\subsection{Predicted Characteristics of \sgrastar\ at 330 
MHz}\label{sec:predict}

The expected size of \sgrastar\  was calculated from the measurements from 
5~to~45~GHz
by Bower et al.~(2003, in preparation).  They find the scattering disk of \sgrastar\ to be
$(1.39\pm 0.02 \times 0.65^{+0.11}_{-0.14})$ milliarcseconds at $\lambda = 1$ cm.  Scaling these dimensions by
$\lambda^2$ as expected from interstellar scattering  gives an expected 330 MHz diameter 
of $(11.5\pm0.2 \times 5.4^{+0.9}_{-1.1})$ arcseconds.  The position angle of the scattering disk of \sgrastar\ is
constant at $\sim 80\arcdeg$ over the range of~1.4 to~46~GHz.  We have
assumed that its orientation remains constant below~$1.4$~GHz.  \sgrastar\ has a 
time averaged flux density at $1.4$~GHz of $\approx 0.5$ Jy and a time averaged 
spectral index  over the centimeter radio regime of $\alpha\sim 0.3$ ($S_{\nu}\propto 
\nu^{\alpha}$) \citep{ZBG}.  Assuming no change in the spectral index below 
$1.4$~GHz, the expected 330 MHz flux density is roughly 0.4 Jy.

\subsection{The Detection of \sgrastar}\label{sec:ia}

\begin{figure}[tbp]
\epsscale{0.9}
\plottwo{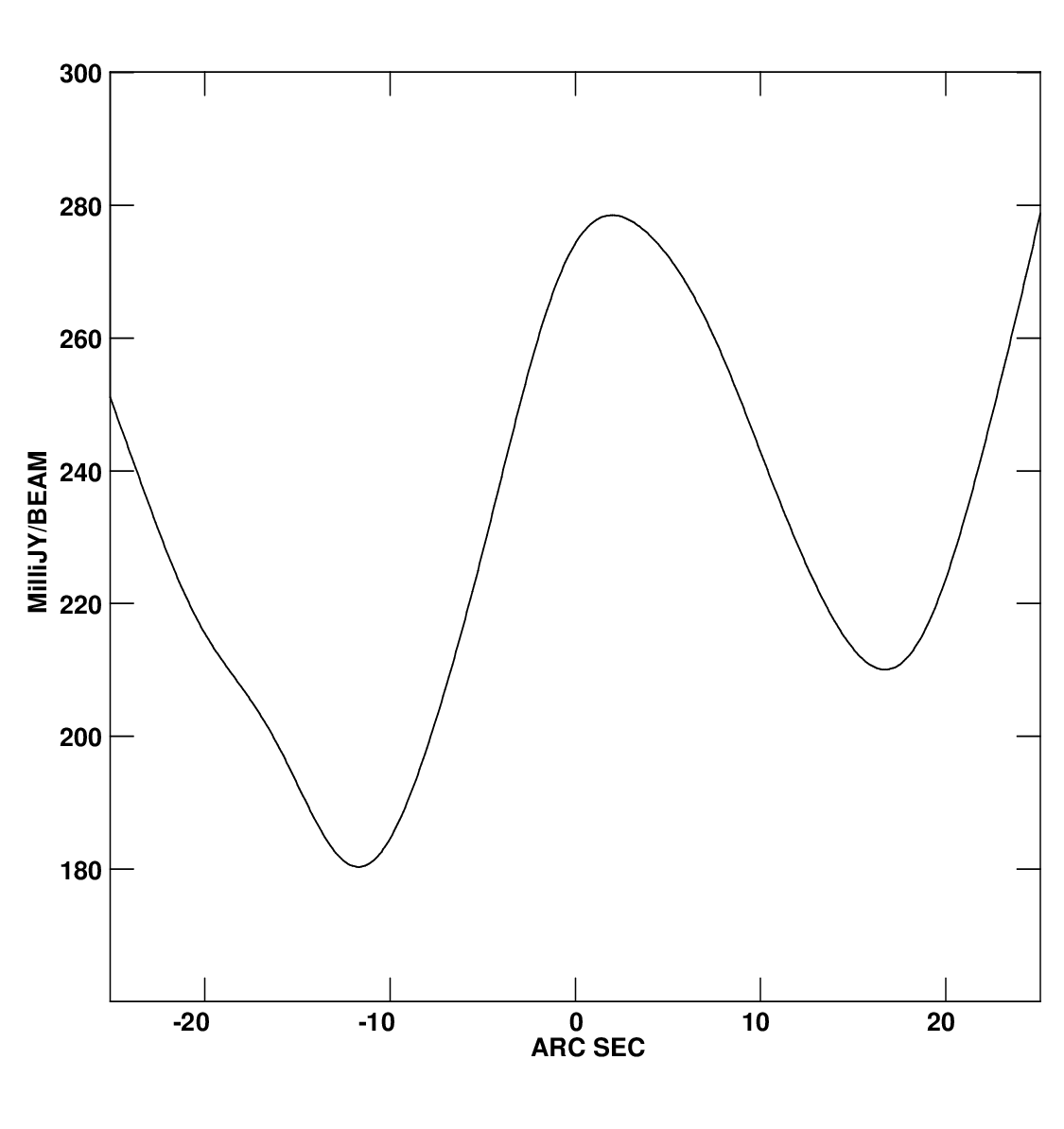}{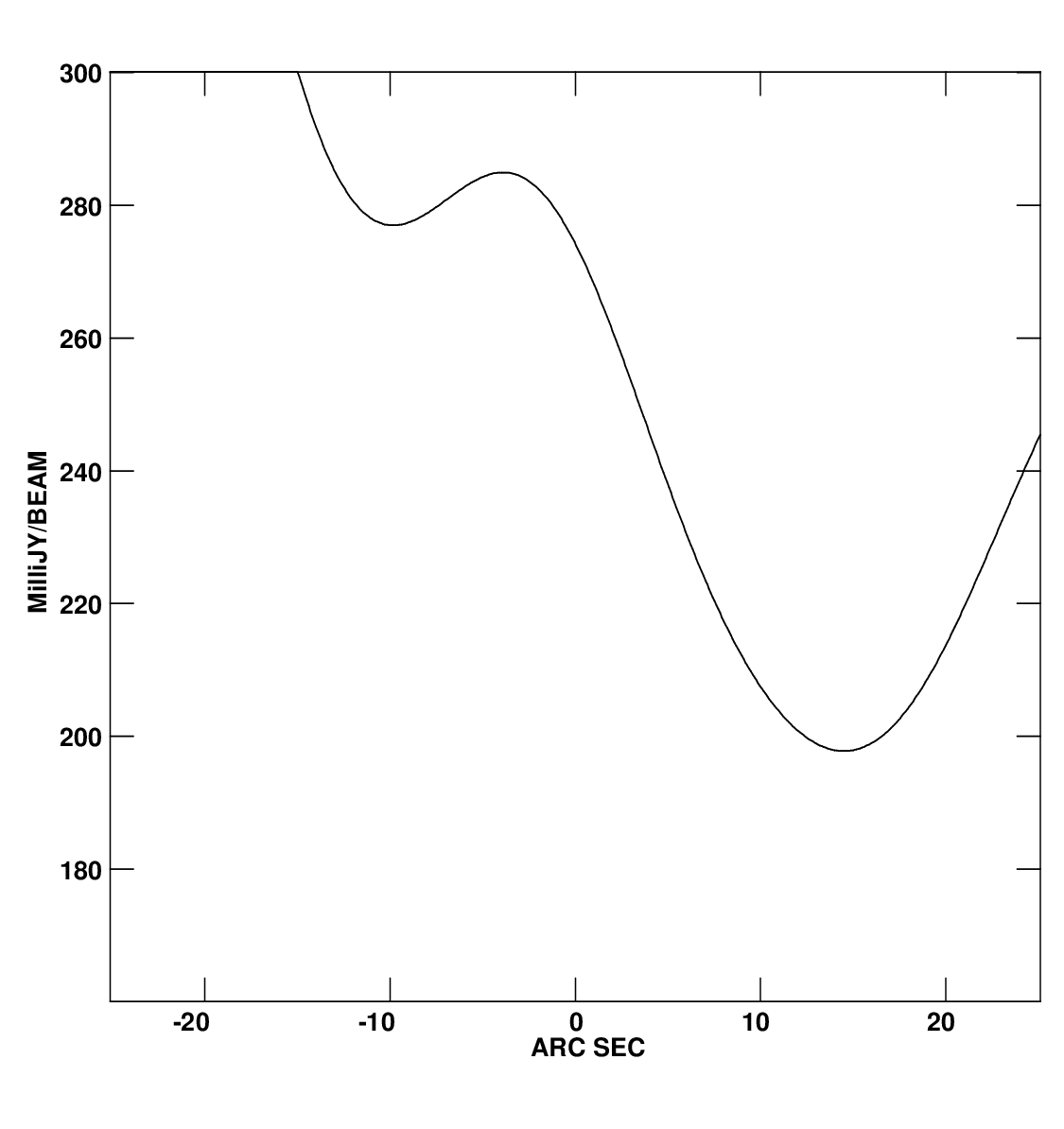}
\caption[]{
\textit{Left} Slice through the expected major axis of \sgrastar\ in
Figure~\ref{fig:sgrastar}.  The slice is centered at 17$^{\mathrm{h}}$
45$^{\mathrm{m}}$ 40$^{\mathrm{s}}$ $-$29\arcdeg 00\arcmin 28\arcsec\
with a position angle of $80\arcdeg$~(J2000) and a resolution of $\sim 6\farcs8$.  East is towards the left.
\textit{Right} Slice through the expected minor axis of \sgrastar\ in 
Figure~\ref{fig:sgrastar}.  Resolution is $\sim 10\farcs9$.  South is towards the left.}
\label{fig:slices}
\end{figure}

Figure \ref{fig:slices} shows slices 
through the assumed major and minor axes of \sgrastar.  
Gaussian fits with baseline subtraction to these slices results in a measured source size 
of $(17\farcs6 \pm 4) \times (15\farcs9 \pm 5)$, a peak intensity of 
80~$\pm$~15~\mjybm, and a total flux density of 330~$\pm$ 120~mJy.  The source deconvolved 
with the $6\farcs8 \times 10\farcs9$ beam gives a source size of $(16\arcsec \pm 4)\times (11\arcsec \pm 5)$.  
If the source is fit in the image instead of fitting slices,, the position angle is then a free parameter instead of assumed to be 80\arcdeg. A 2 dimensional Gaussian fit of the region  yields a position angle of $35\arcdeg \pm 35$.  However, this fit is contaminated by flux density to the south of the source as shown in the slice on the right in Figure \ref{fig:slices}.

The location and flux density of the source agree with the 
extrapolations of \S \ref{sec:predict}, and the size of the source agrees to within $ \sim 1.2 \sigma$ along the assumed minor axis and $\sim 1.1 \sigma$ along the assumed major axis. However, an alternate interpretation of this emission is that it originates from the Sgr A East SNR. 

The absorption in Figure \ref{fig:sgrastar} is caused by the Sgr A West \htwo\ region, first detected in absorption by Pedlar et al.~(1989).
The Sgr A West \htwo\ region is composed of a dense 
region known as the ``mini-spiral'' (Lo \& Claussen 1983, Ekers et al.~1983), 
known to be orbiting \sgrastar\ (Roberts and Goss 1993), and a more diffuse 
ionized region $ \approx 80\arcsec \times 60\arcsec$ in size, known as Sgr A 
West (extended) (Mezger \& Wink 1986; Pedlar et al.~1989).  Both components of 
Sgr A West are detected in thermal (free-free) absorption against the Sgr A East 
SNR with 330 MHz optical depths ($\tau_{330}$) $>1$  across a large portion  of 
the \htwo\ region (Pedlar et al.~1989). The position of \sgrastar\ is offset from the 
mini-spiral (Zhao \& Goss 1998), therefore in order for Sgr~A East to be
absorbed at the position of  \sgrastar\, most of the absorption must
arise from Sgr~A West (extended).  Estimates of the optical depth
along lines of sight near \sgrastar\ based on higher frequency
measurements ($\nu=5$ GHz, Yusef-Zadeh \& Morris~1987) suggest that  330 MHz optical
depths near  \sgrastar\ are $\tau_{330} \sim 10$.
Moreover, any gap through the Sgr~A West \htwo\ region would have to
occur with the size, shape, and location necessary to allow Sgr A East to be detected with the expected size, shape, and flux density of \sgrastar. 
Given the agreement
in observed and predicted source properties,
the likelihood that Sgr~A~East is detected can be discounted.  We conclude 
that we have detected \sgrastar\ at~330~MHz.

\section{Implications}

\subsection{Previous Non-detections}

Previous non-detections at similar frequencies can be attributed to the limited sensitivities
of the observations.  Davies et al.~(1976) observed with the Mk~I-Defford
interferometer.  This two-element interferometer has a north-south
resolution of $\theta \approx 2\farcs4$ at~408~MHz.  
The correlated flux density that would have been measured by this interferometer 
is $S_{\mathrm{corr}} = S_0e^{-(\theta_s/\theta)^2}$, where $\theta_s$ is the 
scattering diameter and $S_0$ is the intrinsic flux density (Roy et al.~2003).  
The expected scattering diameter of \sgrastar\ is 4\farcs1.  Their upper limit, 
$S_{\mathrm{corr}} < 50$~mJy, implies that they would not have detected any 
source weaker than 0.9~Jy, a value that is well above the flux density that we have
observed at 330 MHz.

Pedlar et al.~(1989) observed the Galactic center with the VLA.  At the time 
of their
observations, only a few ($\sim 10$) of the 27 VLA antennas were
outfitted for 330~MHz observations.  In addition to a higher thermal
noise level (rms sensitivity $\approx 16~\mjybm$), the number of baselines 
was substantially lower, resulting in less complete $u$-$v$ coverage.
\sgrastar\ would have been only a 3$\sigma$ source in a crowded and
confusing field.

In 1999, Anantharamaiah, Pedlar, \& Goss~reanalyzed data from Anantharamaiah et 
al.~(1991) in which the GC was observed at 330 MHz with all antennas of the VLA. 
In their Figure 2, there is a hint of a detection of \sgrastar\ at a level of 
$\sim 40$~\mjybm, which is consistent with this detection.  The authors do not comment on this possible detection.

\subsection{The Spectrum of \sgrastar}\label{sec:spectrum}

\begin{figure}[tbp]
\plotone{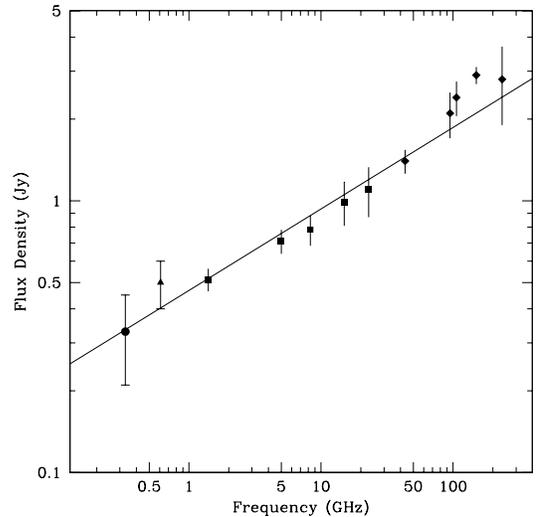}
\caption {The radio/sub-millimeter ($0.33 < \nu < 235$ GHz) spectrum of 
\sgrastar.  Bars without endcaps signify variability limits. The 610 MHz value 
is from Roy et al.~(2003).  The $1.4 < \nu < 23$ GHz values (squares) are from 
Zhao, Bower, \& Goss (2001).  These data are time averaged over 1980-2000. The 
$43 < \nu < 235$ GHz values (diamonds) are from Falcke et al.~(1998) and are 
averaged over 1987-1994. For reference, the solid line indicates a spectrum with 
spectral index $0.3$, it is not a fit to the data.  The low frequency spectral index
is $\alpha^{1400}_{330} = 0.3^{+0.5}_{-0.4}$.}
\label{fig:spectrum}
\end{figure}

Figure \ref{fig:spectrum} shows the observed  spectrum for \sgrastar, including
our flux density measurement.  This spectrum is not 
constructed from simultaneous measurements and as \sgrastar\ is variable, 
simultaneous measurements may show  differences.

There are at least two models of the low frequency spectrum
of \sgrastar\ that are suggested by our detection.  Either the $S_{\nu}
\propto \nu^{0.3}$ spectrum observed across the centimeter radio band
continues toward low frequencies and the source is unobscured by
intervening thermal gas, or the spectrum of \sgrastar\ rises below
$\sim$ 1 GHz and is obscured by significant optical depth.

Assuming a $S_{\nu} \propto \nu^{0.3}$ spectrum, a 330 MHz
optical depth of $\tau_{330} < 0.1$ is required to explain the
measured \sgrastar\ flux density.  Even a flat spectrum below 1.4 GHz
permits an optical depth of only $\tau_{330} \sim 0.4$.  Therefore,
the free-free optical depth along this line of sight is much
less than the $\tau_{330} \sim$ 3--5 previously assumed to explain 
low frequency non-detections (Pedlar et al.~1989).

A conclusion of \S \ref{sec:ia} is that lines of sight to Sgr~A~East that pass near \sgrastar\
have large 330 MHz optical depths.  If the line of sight to \sgrastar\
were similarly obscured, its intrinsic flux density would have to be
increasing at low frequencies in order to agree with the observed flux
density. The optical depth has a frequency dependence of $\nu^{-2.1}$.
In order to agree with the observed flux density, the intrinsic flux
density would have to rise exponentially ($S_{\nu} \propto e^{\nu^{-2.1}}$) below $\sim$ 1 GHz.

Low 330 MHz optical depths near \sgrastar\ are also indicated
by the  22 GHz, $0.1\arcsec$
resolution image of the \sgrastar\ region by Zhao and Goss (1998),
which shows emission measures due to ionized gas within $0.2\arcsec$
of \sgrastar\  are nearly two orders of magnitude lower than in
the nearby mini-spiral.  Furthermore, an exponential
rise in intrinsic flux density to almost exactly cancel
the large (factor of $\sim 10$ between 1000 and 330 MHz) increase in free-free optical
depth at 330 MHz would be quite improbable .

We conclude that it is likely that there is little or no free-free
absorption along the line of sight to \sgrastar, and that the
observed flux density is intrinsic to the source.  Our results imply a
large optical depth towards the Sgr~A East SNR, but a small optical depth
towards \sgrastar.  This could be explained by a localized clearing of the ambient gas accomplished either through the direct influence of the black hole, or through the stellar winds of the central cluster.  It is also possible that the ionized medium in the GC region is
clumped, with the line of sight to \sgrastar\ passing through a
gap.  Our data cannot differentiate between these possibilities.


\subsection{Emission Mechanisms}
Previous modeling of the spectrum of \sgrastar\ did not have to
confront its flux density at low frequencies, as \sgrastar\ was
thought to be unobservable below 1 GHz (e.g.~Mahadevan 1998).  Furthermore, the 410 MHz
non-detection of Davies et al.~(1976) is often incorrectly treated as
an upper limit. Our 330 MHz result and the 610 MHz detection by Roy et
al.~(2003) require these models to be reevaluated.  Because we cannot exclude the possibility of free-free optical depth toward Sgr A*, our reported 330
MHz flux density should be considered a \emph{lower} limit.

Advection dominated accretion flow (ADAF) models (e.g.~Yuan, Markoff \& Falcke 
2002) predict flux densities nearly two orders of magnitude lower than the 330
MHz measurement.  The Relativistic Jet plus ADAF model of Yuan et al.~(2002) 
does not explicitly make predictions below $1.4$~GHz, but would appear to 
closely match the measured 330 MHz flux density.  It should be noted
that in the  jet models, low frequency emission arises from farther out along the jet, probing
regions farther from the event horizon.

\section{Conclusions}

We have presented a new high resolution, high sensitivity image of the
\objectname[]{Sgr~A} region at~330~MHz, constructed from observations at multiple
epochs with the Very Large
Array.  We report a detection of \sgrastar, with a flux density
of $330 \pm 120$~mJy.  Previous non-detections of
\sgrastar\ at comparable frequencies are attributable to the poor sensitivity 
and/or $u$-$v$
coverage of the interferometers used.

We argue that only small amounts of free-free absorption
($\tau_{330} \sim 0.1$) exist along the line of sight to \sgrastar, and
therefore that \sgrastar\ is observed through a region of low density
in the \objectname[]{Sgr~A West} \htwo\ region.  It is possible that
the intrinsic spectrum of \sgrastar\ rises below $\sim 1$ GHz.  However, the intrinsic rise in flux density would have to be balanced by free-free absorption to result in the observed value.  
Therefore, this flux density measurement is a \emph{lower} limit to
the intrinsic flux density of this source as the optical depth along
the line of sight remains unknown (e.g., Anantharamaiah et al.~1999).

This detection at~330~MHz, combined with the recent detection at~610~MHz (Roy et
al.~2003), expands the frequency range over which \sgrastar\ has been detected and allows for 
low frequency observational constraints on \sgrastar\ emission mechanisms.
Future observations at and below~240~MHz with the Giant Metrewave Radio 
Telescope (GMRT) and the Low Frequency Array (LOFAR; Kassim et al.~2001) may 
expand the frequency range at which \sgrastar\ is detectable to even lower frequencies.


\acknowledgements
We thank G.~Bower for the use of unpublished results pertaining to the scattering size of \sgrastar, C.~Brogan for help with data reduction and imaging, and A.~Cohen and 
L.J.~Rickard for helpful discussions.
Basic research in radio astronomy at the NRL is supported by the Office of Naval 
Research.  The National Radio Astronomy Observatory is a facility of the 
National Science Foundation operated under cooperative agreement by Associated 
Universities, Inc.


{}

\clearpage



\end{document}